**Title:**

A random matrix approach to detect defects in a strongly scattering polycrystal: how the memory effect can help overcome multiple scattering.

**Authors:**

S. Shahjahan[1], A. Aubry[2], F. Rupin[1], B. Chassignole[1], A. Derode[2]

**Affiliations:**

[1]EDF R&D - Site des Renardières, 77818 Moret sur Loing, France.

[2]Institut Langevin, ESPCI ParisTech, CNRS, Université Paris Diderot-Paris 7, 1 rue Jussieu, 75005 Paris, France.

**Abstract:**

We report on ultrasonic imaging in a random heterogeneous medium. The goal is to detect flaws embedded deeply into a polycrystalline material. A 64-element array of piezoelectric transmitters/receivers at a central frequency of 5 MHz is used to capture the Green's matrix in a backscattering configuration. Because of multiple scattering, conventional imaging completely fails to detect the deepest flaws. We utilize a random matrix approach, taking advantage of the deterministic coherence of the backscattered wave-field which is characteristic of single scattering and related to the memory effect. This allows us to separate single and multiple scattering contributions. As a consequence, we show that flaws are detected beyond the conventional limit, as if multiple scattering had been overcome.

When imaging an unknown heterogeneous medium with waves (whatever their nature) a fundamental issue is the importance of multiple scattering relative to single scattering. Most imaging techniques rely on a single scattering assumption. Even in the case of more sophisticated imaging approaches, like passive imaging or coda correlation imaging which utilize correlations of multiple-scattered waves to retrieve the Green's function between two passive sensors, it is ultimately the ballistic (i.e., unscattered) part of the estimated Green's function that is used to build an image[1,2,3,4,5,6]. In general, multiple scattering is a nightmare for imaging. Light through fog, sound in a dense forest, and electromagnetic waves in a reverberating building are common examples. Because of multiple scattering there is no longer a direct relation between travel time and depth, which makes echolocation impossible beyond a few scattering mean-free paths, when the wave has lost its coherence.

In recent years the advent of multi-element arrays with controllable emitters/receivers has opened up new perspectives, both experimentally and theoretically. In a linear and time-invariant system with $N$ inputs and $M$ outputs (emitters/receivers), everything can be described by a matrix approach. All relevant information is contained in the $N \times M$ matrix of inter-element impulse responses, or Green's matrix. For applications to imaging in random heterogeneous media, this leads to random matrix theory. In this context, recent academic results proposed a solution based on random matrix properties to overcome multiple scattering by separating the single and multiple scattering contributions in the total Green's matrix[7,8]. From a physical point of view, it was shown to be related to the so-called memory effect in random media optics[9,10,11]. Experimental results were presented with ultrasound waves, but limited to the rather academic situation of a large metallic cylindrical target hidden behind a forest of smaller rods, with a thickness of about 3 times the scattering mean-free path.

The issue we address here is the applicability of this approach, beyond a laboratory experiment, to a real scattering material: a polycrystalline alloy. Moreover, the object to be imaged (a defect in the material) is not *behind* a scattering screen but *within* the host medium, so deep that it is invisible by classical imaging techniques. We present experimental results that confirm the great potential of a random matrix approach in order to image objects within a real multiple scattering medium, beyond a few mean-free paths.

In these experiments the sample under study is a block of polycrystalline steel with dimensions 90×90×280 mm$^3$ (Fig. 1). It is a chromium-nickel-based alloy (Inconel 600) which is commonly found in components of the steam generator and some of the reactor vessels in nuclear power plants, due to its resistance to corrosion at high temperature. The sample has undergone a heating treatment to enlarge its grains. It can be thought of as an arrangement of locally anisotropic grains, with equiaxed and uncorrelated crystallographic orientations so that the material appears isotropic at a macroscopic scale. The grain size distribution has been thoroughly studied with micrographic analyses as well as EBSD (Electron Backscattered Diffraction). The mean grain size was found to be 750 µm (median value 680 µm) with a standard deviation of 400 µm (Fig. 2). The longitudinal wave length around 5 MHz (1.2 mm) is comparable to the grain size. When dealing with wave propagation and imaging in a random medium, whatever the nature of the wave and the complexity of the medium, the scattering mean-free path $\ell_S$ is the essential parameter. In transmission, the intensity of the ballistic (unscattered) part of the wave, the one which is taken advantage of for imaging, decays exponentially with depth z as $exp(-z/\ell_S)$. The scattering contribution from multiple-scattering, which ruins imaging, becomes more and more significant as the penetration depth becomes large compared to $\ell_S$. In the case of polycrystals, knowing the mass density and elastic moduli, one can compute $\ell_S$ at various frequencies, following the "first-order smoothing" (or Keller) approximation to solve Dyson's equation for the ensemble-averaged wavefield[12,13,14,15]. For longitudinal waves, we find $\ell_S \sim$ 22 mm at 5MHz[16], assuming that the microstructural correlation function is exponential with a typical correlation length $\ell_c = 750$ μm, which was verified by EBSD analyses. The scattering mean free path $\ell_S$ is found to decrease with frequency: we obtain $\ell_S \sim$ 36 mm at 3.3 MHz and $\ell_S \sim$ 17 mm at 6.8 MHz, which are the limits of the frequency band. Thus multiple scattering can be expected to play a very significant role around 5MHz, considering the 90 mm-depth which is significantly larger than $\ell_S$. This has been confirmed experimentally by a recent study of the coherent backscattering effect in the same material[17].

One part of the sample was left unchanged (healthy area), while in the other part several artificial defects were machined (Fig. 1). In particular, a series of 1-mm radius cylindrical holes were drilled at depths ranging between 10 and 70 mm, with a 20 mm step. The goal is to detect these defects as deep as possible, despite the strongly scattering structure that surrounds them. Ultimately, for practical applications, nothing can replace real defects. The defects we study here, even if they are artificial, are the first steps towards a real defect such as a fatigue crack. Besides, their cylindrical shape makes them invariant under translation along the y-axis (Fig. 1). Hence it is possible, by translating the array along y, to study the same defect buried under different arrangements of steel grains. Here, eleven independent sets of data were recorded for each flaw.

We used a 64-element array, at a 5.2 MHz central frequency (-6 dB bandwidth 3.3 to 6.8 MHz). The element sizes were 0.3×15 mm² and the pitch was 0.5 mm. The first step consists in acquiring the Green's matrix $G$. To that end, the $i$-th element of the array is used as an impulsive source, and the backscattered waves are recorded on all array elements. The actual input signal is not a Dirac pulse, but a sine wave centered at 5.2 MHz, with a Gaussian envelope exp(-t/2τ²), with τ=0.1μs, which fits the bandpass of the transducer. The procedure is repeated for all values of $i$, from 1 to 64, which yields a series of 64×64 time signals $g_{ij}(t)$, where $j$ is the receiver index. This procedure is sometimes referred to as "full matrix capture". Once the matrix $G$, which contains the inter-elements Green's functions, is known, the rest is only post-processing to manipulate the backscattered field and extract relevant information.

Focused beamforming is a classical way to form an image of the inspected medium from the time signals $g_{ij}(t)$; for instance this is commonly applied in standard medical ultrasound imaging systems[18,19]. Basically the idea is to mimic a lens that would focus ultrasound in emission and reception at a desired focal point with coordinates $(x, z)$. This can be easily done once the Green's matrix has been recorded. Assuming that the medium is a cloud of small scatterers embedded in a medium with a constant sound velocity $c_0$, it is straightforward to convert the difference of distances between each array element and the desired focal point into a set of time-delays. Once the delayed signals are summed, the resulting amplitude at the focal time is kept, to obtain a value representative of the reflectivity of the medium at $(x, z)$. From here on, we will refer to this as TFM (total focusing method)[19]. TFM is an efficient and simple imaging technique but its essential assumption is that multiple scattering must be negligible, so that there is a direct correspondence between time and depth. When multiple scattering becomes too strong, TFM completely fails, as all classical imaging techniques. This is illustrated in Fig. 3, where the defect is no longer detected when it is too deep compared to the scattered mean-free path.

In recent works, an alternative imaging technique based on random matrix theory was proposed[7,8]. Though it is out of the scope of this letter to explain it completely, we briefly recall the underlying ideas. In any medium, the received waves contain both single (SS) and multiple scattering (MS) contributions, so that the Green's matrix may be written as $G = G_{SS} + G_{MS}$. Here a Fourier transform, or even a full time-frequency analysis, has been performed on $g_{ij}$, so that $G$ is a complex-valued $N \times N$ matrix; $N$ being the number of array elements. The time and frequency dependence of $G$, $G_{SS}$ and $G_{MS}$ have been omitted for simplicity.

At early times (i.e., depths smaller than or comparable to the scattering mean-free path $\ell_S$) SS dominates and classical imaging works efficiently. At later times, multiple scattering is predominant, beamforming fails, and ultimately propagation becomes entirely diffuse[20]. At intermediate times, single and multiple scattering contributions coexist. Interestingly, it was shown that though $G_{SS}$ and $G_{MS}$ are both random matrices, they do not have the same statistical behavior at all. In particular, while $G_{MS}$ has only a short-range spatial correlation, $G_{SS}$ displays a long-range correlation along its antidiagonals (i.e., for array elements $i$ and $j$ such that $i + j$ is constant), whatever the realization of disorder (for an illustration, see for instance Fig. 3 in Ref. 8). Physically, this can be understood as the equivalent, in a backscattering configuration, of the well-known "memory effect" in optics[10,11]. When an incident plane wave is rotated by an angle $\theta$, the far field speckle image measured in transmission is shifted by the same angle $\theta$ (or $-\theta$ if the measurement is carried out in reflection), as long as $\theta$ does not exceed the angular correlation width $\Delta\theta$. In the single scattering regime, $\Delta\theta = \pi/2$, hence the memory effect spreads over the whole angular spectrum[10,11]. This accounts for the fact that the matrix coefficients are coherent along a given antidiagonal when only single scattering takes place. Indeed if two pairs of array elements $(i_1; j_1)$ and $(i_2; j_2)$ are on the same antidiagonal then we have $i_1 + j_1 = i_2 + j_2$. Changing the direction of emission amounts to changing $i_1$ into $i_2$. As a result, in reflection the speckle image will be tilted so that the signal that was received in $j_1$ will be coherent with the new signal in $j_2 = j_1 - (i_2 - i_1)$. On the contrary, when MS dominates, the correlation width $\Delta\theta$ is much smaller (in the diffuse regime it diminishes as $t^{-1/2}$) and as soon as $\Delta\theta$ becomes smaller than the angular aperture of one array element, the matrix coefficients become random and uncorrelated (except for the fact that reciprocity implies that $G$ is always symmetric)[21].

This can be taken advantage of to separate the single-scattering and the multiple-scattering contributions. This is achieved via rotations and projection of the Green's matrix (technical details are given in Refs 8 and 22). These manipulations amount to selecting, within $G$, the part of it that presents the aforementioned "antidiagonal correlation". The resulting "filtered" matrix will be denoted $G^F$. Hopefully it corresponds to the single scattering contribution $G_{SS}$, albeit weak. It is thus a way to get rid of multiple scattering, thereby improving imaging.

Finally, once the SS/MS separation is performed, an image of the medium can be built from $G^F$, based on a singular value decomposition (SVD) and the DORT method[23,24]. We will term this MSF-DORT, MSF meaning "multiple scattering filter". Let us recall that SVD consists of factorizing any matrix $M$ as $M = U\Lambda V^*$, where $U$ and $V$ are unitary matrices containing singular vectors while $\Lambda$ is a diagonal matrix whose nonzero elements $\lambda_i$ are the singular values. They are always real and positive. The basic idea underlying DORT (the French acronym for decomposition of the time-reversal operator) is the following: in the single-scattering regime and for point-like scatterers, each scatterer is associated with only one significant singular vector of the Green's matrix $G$, corresponding to a nonzero singular value of $G$. Physically, each singular vector corresponds to the wave front that, if it was sent from the array, would focus onto the corresponding scatterer; and the associated singular value $\lambda_i$ is proportional to its reflectivity. Therefore, by numerically back-propagating a given singular vector in the supposedly homogeneous medium, an image of the corresponding scatterer can be obtained. In the imaging experiment presented here, we try to detect a defect embedded in a scattering structure. We assume that, if a defect is present and multiple scattering can be removed, then its echo will correspond to the largest of the singular values (i.e. $\lambda_1^F$) of the "filtered" matrix $G^F$ and the corresponding singular vector $V_1^F$. Therefore, an image is built by back-propagating the first singular vector as: $\lambda_1|V_1 G_0|$, where $G_0$ is the Green's matrix in a supposedly homogeneous reference medium. Here, for all imaging methods, we considered a constant velocity $c_0 =$5850 m/s, corresponding to the effective longitudinal velocity that was measured in the polycrystalline alloy.

The resulting images are presented in Fig. 4. They show a dramatic improvement, since the deepest flaws (50 mm and 70 mm) appear to be clearly detected, compared to Fig. 3. However, one has to go beyond a merely graphical comparison, and examine on what grounds a defect is said to be detected or not. This requires a finer analysis, given that the quantities at stake in TFM and MSF-DORT are essentially very different. But what both approaches have in common (at least in random scattering media) is that they both deal with random variables. In the case of DORT, the relevant random variable is $\lambda_1$. In the case of TFM, it is the brightness of the image after beamforming. In order to establish an unbiased comparison, we used a criterion based on the probability of false alarm (PFA). Let us denote by $X$ the physical quantity that is used to decide whether a defect is detected or not. We start by studying the healthy area, whose statistical characteristics are identical to those of the flawed area. Using every face of the sample and as many positions as possible we acquire enough experimental data to have consistent statistical information about the Green matrix $G$. Hence we can estimate the probability density function of $X$ ($f_X$) as well as the cumulative distribution function ($F_X$) for any random variable $X$ related to $G$. Once $F_X$ is known, we set a probability of false alarm (for instance PFA=1%) and deduce the corresponding threshold $\alpha = F_X^{-1}(1 - PFA)$. In other words, if in later experiments we observe a value of $X$ larger than $\alpha$, we will consider this value as abnormal and will conclude that a defect is present, with a 1% chance of being wrong. If $X$ is smaller than $\alpha$ in the area where the defect is supposed to be, then it is not detected. This procedure enables us to compare various imaging techniques objectively, on a common ground. In Fig. 4, we have set the PFA at the same level 1%, and applied the corresponding thresholds to the TFM and MSF-DORT techniques.

The results in Fig. 4 are impressive, but they were obtained on a single realization of disorder. Given that the microstructure is randomly changing, one might think that this very good result is just a lucky strike. To address this issue, we have translated the array along the axis of the cylindrical holes (axis y in Fig. 1) and repeated the experiment. This gave us 11 independent realizations of the microstructure for the same defect. The 70 mm-deep defect was never detected with TFM, whereas it was detected 8 times out of 11 once the "multiple scattering filter" was applied. Although 11 is not a huge number, it is enough to conclude that the improvement brought by the separation of single and multiple scattering contributions yields a statistically significant improvement.

The experimental results presented here clearly demonstrate the efficiency of the SS/MS separation based on a matrix approach, in the context of ultrasonic imaging in strongly scattering materials. In the polycrystalline alloy studied here the probability of detection of a given defect embedded at about 3 times the mean-free path was shown to be dramatically improved, from 0% to almost 73%. Further theoretical and experimental work is needed, for instance in order to model the relation between microstructure (grain size distribution, elastic moduli) and statistical parameters of the Green's matrix (probability density function of the singular values), and to account for more complicated flaws like fatigue fractures or stress corrosion cracks.

**Acknowledgments**

A. A. and A. D. are grateful for funding provided by the Agence Nationale de la Recherche (ANR-11-BS09-007-01, Research Project DiAMAN), LABEX WIFI (Laboratory of Excellence ANR-10-LABX-24) within the French Program "Investments for the Future" under reference ANR-10-IDEX-0001-02 PSL*, and by Electricité de France R&D. All authors thank Dr. David Andrew Smith for reading and editing the manuscript.

**Figure 1**

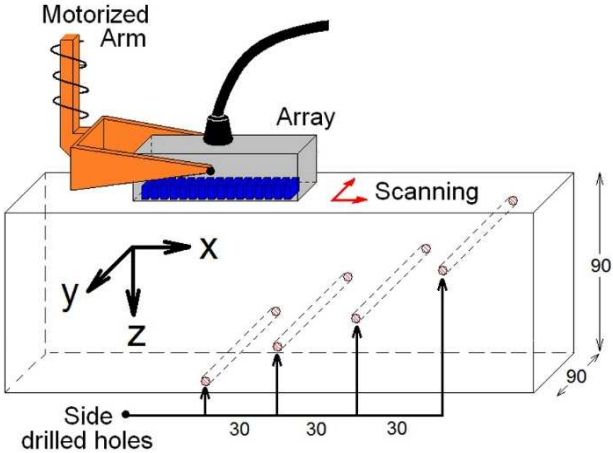

Fig. 1: Experimental set-up. The 64-element array is placed in contact with the material. Its position is controlled by a motorized arm. In this sketch, the left part of the sample is the healthy area (no flaws). In the right part, four cylindrical holes have been drilled at 10, 30, 50 and 70 mm from the top face (i.e. 20, 40, 60 and 80 mm from the bottom face). The array can be placed on any face of the sample.

**Figure 2**

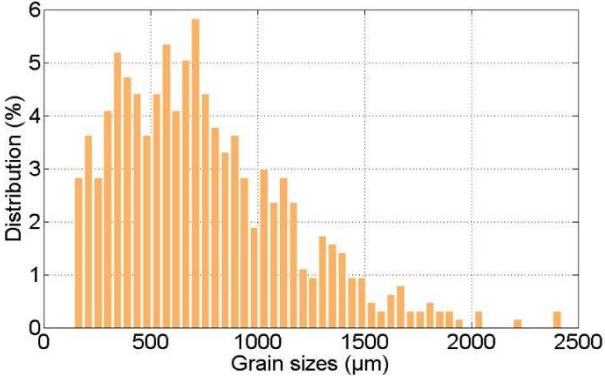

Fig. 2: Microstructure of the scattering sample. A micrograph obtained after electro-etching reveals the grain boundaries. A statistical analysis of the resulting image yields the grain size distribution.

**Figure 3**

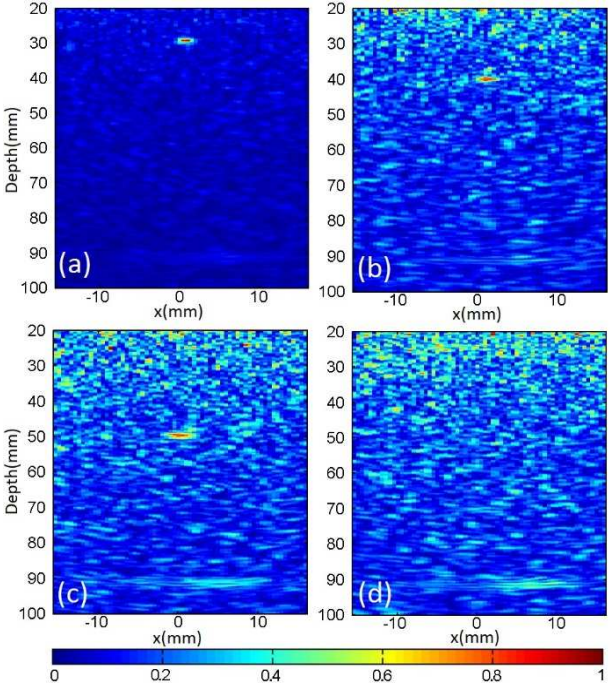

Fig. 3: Conventional TFM imaging of flaws at depth 30 mm (a), 40 mm (b), 50 mm (c) and 70 mm (d). The flaw is well detected at 30 mm depth, barely visible at 50 mm and completely overwhelmed by multiple scattering at 70 mm. In these pictures, the center of the array coincides with the defect position (x=0). Note that the images of the cylinder cross sections appear elongated because of the different axes scales. Each picture is normalized by its maximum.

**Figure 4**

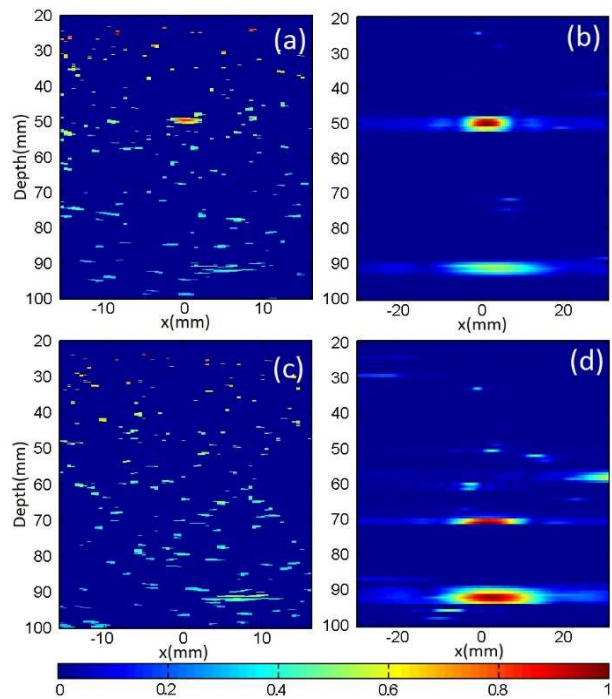

Fig. 4: Comparison between images obtained with or without the "multiple scattering filter". At each time t (corresponding depth $c_0 t/2$) the thresholds $\alpha$ are determined by the probability of false alarm, which is set at 1% for both techniques. TFM images are on the left, MSF-DORT on the right. The flaw depth is 50 mm (a and b) and 70 mm (c and d). Both defects are very clearly detected once the multiple scattering filter is applied (b and d). Also note that the backwall echo, around z=90 mm, is much more visible. Each picture is normalized by its maximum.